\documentclass{article}
\usepackage{ijcai22}

\usepackage{times}
\usepackage{soul}
\usepackage{url}
\usepackage[hidelinks]{hyperref}
\usepackage[utf8]{inputenc}
\usepackage[small]{caption}
\usepackage{graphicx}
\usepackage{amsmath}
\usepackage{amsthm}
\usepackage{booktabs}
\usepackage{algorithm}
\usepackage{algorithmic}
\usepackage{multirow}
\usepackage{bm}
\usepackage{amssymb}
\usepackage{amsfonts} 
\usepackage{nicefrac} 
\usepackage[dvipsnames]{xcolor}
\newcommand{\aticks}{a^{\textrm{ticks}}}
\newcommand{\apct}{a^{\textrm{pct}}}

\title{Learn Continuously, Act Discretely:\\ Hybrid Action-Space Reinforcement Learning For Optimal Execution}

\author{
Feiyang Pan$^{*1}$
\and
Tongzhe Zhang$^{*2}$
\and
Ling Luo$^1$
\and
Jia He$^1$
\And
Shuoling Liu$^2$\\
\affiliations
$^1$Huawei EI Innovation Lab\\
$^2$E Fund Innovation Lab\\
\emails
pfy824@gmail.com,
ztz1992@foxmail.com
}
\begin{document}

\maketitle

\begin{abstract}
Optimal execution is a sequential decision-making problem for cost-saving in algorithmic trading. Studies have found that reinforcement learning (RL) can help decide the order-splitting sizes. However, a problem remains unsolved: how to place limit orders at appropriate limit prices?

The key challenge lies in the ``continuous-discrete duality'' of the action space. On the one hand, the continuous action space using percentage changes in prices is preferred for generalization. On the other hand, the trader eventually needs to choose limit prices discretely due to the existence of the tick size, which requires specialization for every single stock with different characteristics (e.g., the liquidity and the price range). 
So we need continuous control for generalization and discrete control for specialization. To this end, we propose a hybrid RL method to combine the advantages of both of them. We first use a continuous control agent to scope an action subset, then deploy a fine-grained agent to choose a specific limit price. Extensive experiments show that our method has higher sample efficiency and better training stability than existing RL algorithms and significantly outperforms previous learning-based methods for order execution.
\end{abstract}

\section{Introduction}

Since computer algorithms encompass the whole trading process in modern financial markets, optimal execution has attracted significant research attention for decades. For the buy-side, optimal execution aims to save cost and reduce market impact when executing large orders. It is often viewed as a stochastic sequential decision-making process because large orders need to be divided into a series of small sub-orders within a period of time to be filled without influencing the market too much. Traders or algorithmic trading systems should decide the prices and volumes of the split sub-orders during this process.

There are two fundamental types of transactions: market orders and limit orders. Market orders are transactions meant to be executed as quickly as possible at the market price. Therefore, when using market orders, the agent only decides the size of each sub-order. Many studies showed that modern RL is powerful for such a problem \cite{nevmyvaka2006reinforcement,hendricks2014reinforcement,ning2018double,fang2021universal}.

However, there are two disadvantages for such methods: 1) market orders have no control over the price and might be filled at disappointing prices if the bid-ask spreads are wide, and 2) the volume of sub-orders given by RL agents might fluctuate too much because they work as a sort of market timing, potentially causing large market impact. So in this paper, in contrast to previous work, we focus on optimal execution with limit orders rather than market orders.

A limit order is an order to buy or sell a stock at a specific price or better, which can only be filled if the stock's market price reaches the limit price. Therefore, an agent for optimal limit order placement should control both limit prices and volumes of sub-orders. 

A challenge emerges when using a learning-based agent to decide the limit price: the action space can be viewed as both discrete and continuous. From the perspective of the original problem, the choices of limit prices are discrete due to the existence of the tick size (the minimum price increment). For example, for a \$10.00 stock, one may set bid prices at \$10.00 or \$9.99, but not at \$9.995. So an ideal end-to-end agent should use discrete control to select the actual prices precisely, but it is usually impossible to generalize between stocks with different price-level and liquidity. For example, an action of $-1$ tick means $-1\%$ for a \$1.00 stock but $-0.01\%$ for a \$100.00 stock. 

On the other hand, people usually use percentages to understand the rise and fall of stock prices, which is a continuous representation. As training deep RL agents with high-dimensional and noisy state dynamics requires a lot of trial-and-error, continuous control that determines the percentage change is preferred for its generalization ability across stocks with different prices. For example, a bid at -0.1\% results in \$9.99 for a \$10.00 stock, and results in \$99.90 for a \$100.00 stock. However, such a method is more inclined to learn the strategy of stocks with high prices and liquidity, and fails to specialize in determining the ticks for illiquid low-priced stocks.

In light of these observations, we propose Hybrid Action-space Limit Order Placement (HALOP), which trains an agent to learn continuously and act discretely. First, for generalization, HALOP employs a continuous-control agent to scope an action subset with high-dimensional market dynamics as input. Next, a fine-grained discrete-control agent chooses a specific discrete limit price from the action subset, which specializes well in stocks with different characteristics. Both agents are trained end-to-end to maximize the long-term return, i.e., the excess return in execution. 

We conduct extensive experiments and show that our new optimal execution method with limit orders can beat the market, and significantly improves the excess return upon previous learning-based order execution methods.

\section{Related Work}

\subsection{Optimal Execution}

\subsubsection{Non-machine-learning Order Execution}
Traditional optimal execution research tends to assume that the market price movements follow some stochastic process such as the Brownian motion, and then use stochastic optimal control method to derive the volume trajectory analytically~\cite{almgren2001optimal,2000Optimal,bertsimas1998optimal}. 
However, practitioners rarely use these methods to place orders in stock markets, because the assumptions may not hold in the real-world. The most widely used trade execution strategies are based on pure rules or statistical rules. For example, the time-weighted average price (TWAP)~\cite{bertsimas1998optimal} strategy divides a large order into equal-sized sub-orders and executes each within equally divided time intervals.

The volume-weighted average price (VWAP)~\cite{kakade2004competitive} strategy first estimates the average volume traded for each time interval from historical data, then splits order based on the estimates.

Although simple enough, TWAP and VWAP are still popular these days because their execution costs are always close to the market. Specifically, we let the TWAP strategy with market order be the benchmark in this paper.

\subsubsection{Reinforcement Learning for Order Execution} 
% Since the use of RL in finance has attracted attention for decades~\cite{1998Reinforcement}, many have found that the application of RL for order execution is a perfect match. 
As order execution is fundamentally a problem of making decisions under uncertainty and the actual trading data is full of noise, RL becomes the best choice to solve such a problem.
Nevmyvaka \cite{nevmyvaka2006reinforcement} is the pioneer for RL in optimal execution where the agent is trained by Q-Learning~\cite{1992Q} to chooses the limit price. However, they only considers a small limited set of discrete actions (a few bid and ask prices) without considering the more important percent changes in actions.

With the development of deep learning and deep RL in the past few years, some studies use deep RL to learn to execute orders with high-dimensional market data as inputs
\cite{hendricks2014reinforcement,ning2018double,lin2019optimal,lin2020end,fang2021universal}.
While \cite{hendricks2014reinforcement} applies RL to modify a given volume trajectory suggested by the traditional order execution model Almgren-Chriss~\cite{hendricks2014reinforcement}.
Without any market assumptions, several variations of Deep Q-Network (DQN)~\cite{mnih2013atari} are proposed to choose discrete volumes, which could address the high dimensions and the complexity of the finance market and trading signals with the deep neural network~\cite{ning2018double,lin2019optimal}.
Instead of manually designed attributes, a PPO-based optimal execution framework is designed to make decisions based on raw level-2 market data~\cite{lin2020end}.
Policy distillation paradigm is also deployed in order execution \cite{fang2021universal}, which uses a distilled PPO agent to choose optimal order-splitting volumes.
However, most of the previous work studies the problem of volume-splitting, which only addresses the order execution problem partially. 

\section{Hybrid Action-Space Optimal Execution}

\subsection{Problem Setting}
In this part, we first describe the general setting of the optimal execution problem with limit orders, then formulate it as a sequential Markov Decision Process (MDP). In particular, we focus on setting the limit prices of sub-orders rather than the volumes for clarity, although the proposed method can work seamlessly together with previous volume-oriented methods.

\subsubsection{General Problem Description} We simplify the problem as a discrete-time decision process following \cite{cartea2015algorithmic,ning2018double,fang2021universal}, where the agent interacts with the environment multiple times within a predefined time period based on discrete time space.
So consider a horizon of $T$ timesteps $t=1,\dots,T$, each represents a $1/T$ of a predefined total time periods. For example, if there is a total time period of three hours and $T=90$, it has $90$ small time periods of two minutes. 
At the $t$-th step (the beginning of the $t$-th small time period), the environment reveals the recent market dynamics to the agent, consisting of a series of history quotes and top 5 bid/ask prices and volumes. Then, the agent sends a limit order with a limit price rounded to the tick size and a volume. Next, at the end of the $1/T$ period, the environment tells the agent whether (or how much) the order has been filled, and withdraw the rest of it. Then it moves to the $(t+1)$-th timestep. Specifically, at the end of the $T$-th period, if there is still inventory remained, the system automatically send a market order to execute it at once.
\begin{figure}[t]
    \centering
    \includegraphics[width=0.48\textwidth]{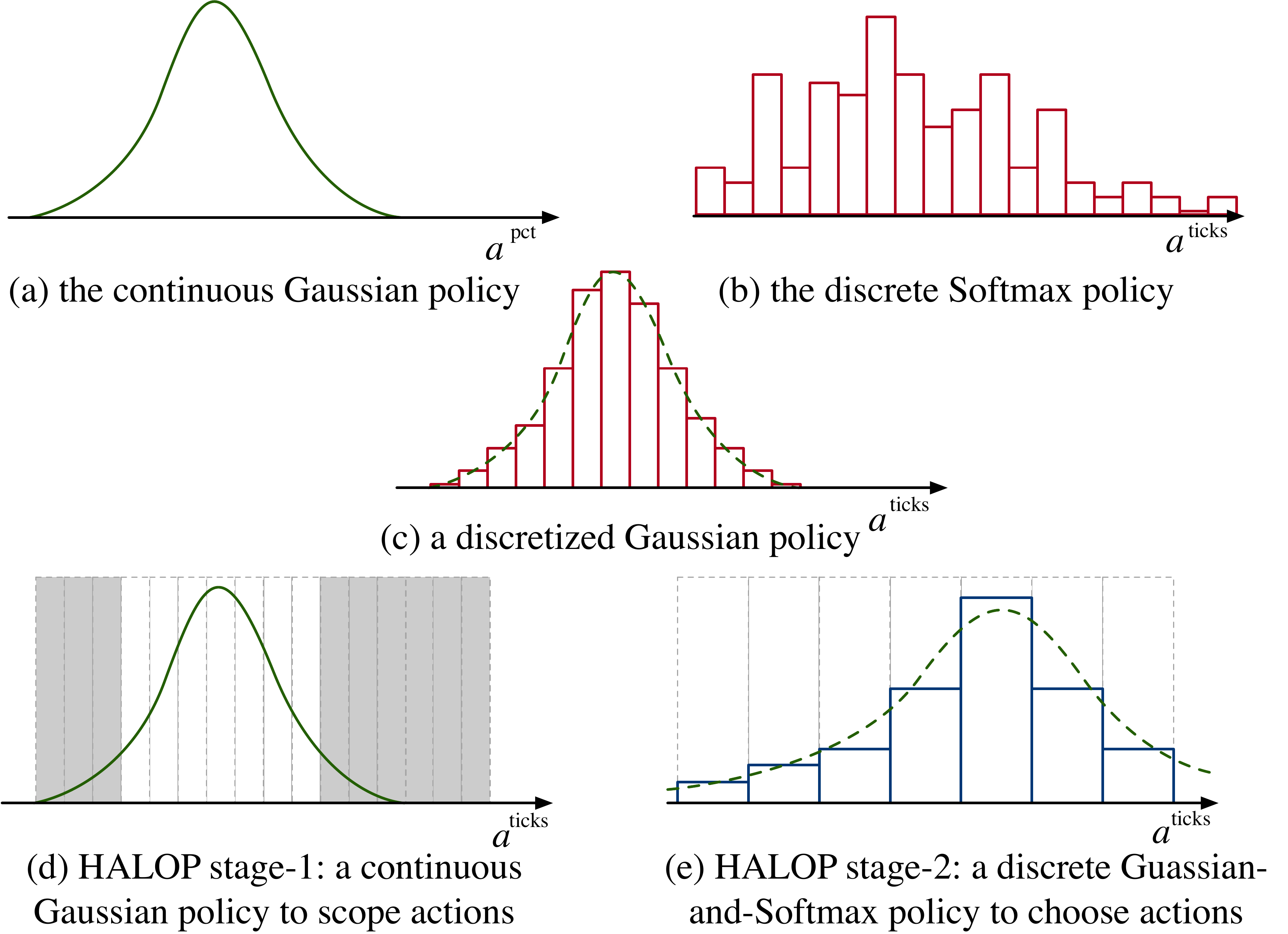}
    \caption{Illustration of different types of policies for different types of action-spaces.}
    \label{fig:actionspace}
\end{figure}

In addition, as we focus on the price, we assume there is a predefined \textit{schedule} of trading volume for the agent, denoted by $v^*_1, \dots, v^*_{T}$, which sums to one $\sum_{t=1}^{T} v^*_t=1$. For example, it can be the volumes of classic financial strategies like TWAP or VWAP, or the outputs of any volume-oriented algorithmic trading agent \cite{fang2021universal}.
As the limit orders are not guaranteed to be executed, let $\tilde v_1, \dots, \tilde v_T$ be the \textit{executed} volume within each period, which has $\Delta_t = \sum_{\tau=1}^{t}v^*_\tau - \sum_{\tau=1}^{t}\tilde v_\tau \geq 0$. So at step $t$, we assume that the agent tries to catch up with the schedule, by placing a new order with a volume 
\begin{equation}
    v_t = v^*_t + \Delta_{t-1}.
\label{eq:vt-def}\end{equation}

\subsubsection{MDP Formulations for Limit Order Placement} Now we formulate the problem as a finite-time MDP, and put forward the continuous-discrete duality of the problem. 

The state consists of two types of information, the public state and the private state $s_t=(s^\textrm{pub}_t, s^\textrm{priv}_t)$. Similar to \cite{nevmyvaka2006reinforcement,lin2020end,fang2021universal}, the public state includes past market dynamics obtained from the environment. Private observations are real-time status of the running agent, including the remaining inventory $1-\sum_{\tau=1}^{t-1}\tilde v_\tau$; the amount of unfulfilled orders $\Delta_{t-1}$; and the remaining time fraction $1-t/T$. 

The action, by default, should be an actual price rounded to the tick size. For example, for a \$10.00 stock, the action of a bid should be like \$9.97. However, such an action changing over time cannot directly be used to formulate an MDP. So we put forward two types of action representations:
\begin{itemize}
    \item Action measured by ticks $\aticks_t \in \mathcal{A}^{\textrm{ticks}} \subset \mathbb{Z} $, i.e., $$\aticks_t=(\textrm{LimitPrice}_t - \textrm{CurrentPrice}_t) / \textrm{TickSize} $$ 
    It is natural to represent price changes in ticks based on the current price\footnote{Here ``current price'' refers to the latest executed price.}. So it is a discrete number, e.g., a bid at \$9.97 for a \$10.00 can be represented as $\aticks=-3$, if the tick size is one cent. 
    \item Action measured by percentage $\apct_t \in \mathcal{A}^{\textrm{pct}} \subset \mathbb{R} $, i.e., $$\apct_t=(\textrm{LimitPrice}_t / \textrm{CurrentPrice}_t-1)\times 100\%$$ It is more general and easier to understand when measuring price movements by percentages or basis points. So any action can has a continuous representation by basis points, which is a continuous variable. E.g., the same bid can be represented as $\apct=-0.3\%$.
\end{itemize}
So apparently one can calculate $\apct$ from any $\aticks$, 
%$$\apct = \frac{\aticks \times \textrm{TickSize}} {\textrm{CurrentPrice}} \times 100\%,$$
but not from the opposite direction if without rounding.
Note that although here we only discuss the limit price, one can always append the volume $v_t$ as the second term of the action.

Finally, we design the reward feedback. Like in real trading systems, we give the agent a reward only at the end of the last period. With the order schedule and the execution status of $t=1,\dots,T$, the reward is
\begin{equation}
R = \mathrm{D} \cdot \bigg[\sum_{t=1}^T v^*_t p^*_t - \bigg(\sum_{t=1}^T \tilde v_t \tilde p_t + \Delta_T \tilde p_{-1}\bigg)\bigg],
\end{equation}
with $p_t^*$ the market price of buying $v^*_t$ in the $t$-th period, $\tilde p_t$ the agent's average execution price, $\Delta_T$ the final remained inventory, $\tilde p_{-1}$ the execution price of the final market order. The multiplier $\mathrm{D}$ denotes the trading direction, $\mathrm{D}=1$ for buying and $\mathrm{D}=-1$ for selling. So using market orders yields a reward of 0.

\subsection{Policy Optimization with Hybrid Action-spaces}
In this part, we put forward our policy optimization method that balances generalization with continuous control and specialization with discrete control. We use a general stochastic policy optimization method, PPO \cite{DBLP:journals/corr/SchulmanWDRK17}, as our base learner. 

There are two basic types of action-spaces, continuous and discrete, and the corresponding stochastic policies, as shown in Figure \ref{fig:actionspace}(a)(b). For the continuous action space $\mathcal{A}^{\textrm{pct}}$ to choose percentage changes in price, one can use a common parametric Gaussian policy with its mean $\mu$ and scale $\sigma$ learned with a policy neural network, i.e.,
\begin{equation*}
    \apct_t|s_t \sim \mathcal{N}(\mu(s_t), \sigma(s_t)).
\end{equation*} 
For the discrete action space $\mathcal{A}^{\textrm{ticks}}$, a common choice is the Softmax policy, which randomly choose actions from a multinoulli categorical distribution
\begin{equation*}
    \aticks_t | s_t \sim \textrm{Categorical}\big\{\textrm{Softmax}(\bm{f}_{1:m}(s_t))\big\},
\end{equation*}
where $\bm{f}_{1:m}(s_t)$ denotes the policy network's output logits for $m$ candidate actions, and $\textrm{Categorical}\{\cdot\}$ is the categorical (multinoulli) distribution. 

Although straight-forward, these policies are hard to optimize because the environment's state dynamics are too complex, the rewards are noisy and sparse, and the action space is too large especially for the discrete policy. To further restrict the search space, we propose a new family of policies: the \emph{discretized Gaussian} policy, and its simplified version, the \emph{Gaussian-and-Softmax} policy.

\subsubsection{Discretized Gaussian Policies}
Consider that we have a policy network that outputs the mean $\mu$ and scale $\sigma$ of some Gaussian distribution $\mathcal{N}$, and a increasing series of predefined \textit{locations} (candidate actions) $a^*_1,\dots, a^*_m$ for discretization. 

Let $a^*_0=-\infty$ and $a^*_{m+1}=+\infty$, for $1\leq k \leq m$
\begin{align}
    d_k &= \textrm{P}(\frac{a^*_{k-1}+a^*_{k}}{2} < a < \frac{a^*_k+a^*_{k+1}}{2}) \\
    &=\Phi(\frac{a^*_k+a^*_{k+1}-2\mu}{2\sigma}) - \Phi(\frac{a^*_{k-1}+a^*_{k}-2\mu}{2\sigma}) \label{eq:disc-gaussian-int}
\end{align}
where $\Phi(x)= \frac{1}{\sqrt{2 \pi}} \int_{-\infty}^{x}\exp\{-\frac{u^2}{2}\} du$ is the CDF of the standard normal distribution. 

So now we can build a discretized Gaussian policy by using a categorical distribution to choose the percentage action, i.e.,
\begin{equation}
    \apct_t|s_t \sim \textrm{Categorical}\big\{d_1, \cdots, d_m \big\}
\end{equation}
where the $a^*$s for calculating $d$s can be determined by the actual tick actions. In this case, such a policy can take the place of the original continuous Gaussian polices.

However, Eq.(\ref{eq:disc-gaussian-int}) involves an integral which is intractable for a feed-forward neural network. Therefore, we use uniform samples from $U(\frac{a_{k-1}+a_{k}}{2},\frac{a_{k}+a_{k+1}}{2})$ and calculate the averaged Gaussian density so that the output action probability can be differentiable. I.e.,
\begin{align}
    \hat d_k = \frac{a_{k+1}-a_{k-1}}{2}&\hat{\mathbb{E}}\bigg[\frac{1}{\sigma\sqrt{2 \pi}} \exp\left\{-\frac{(a-\mu)^2}{2\sigma^2}\right\}\notag \\
    &\big|\, a\sim U(\frac{a_{k-1}+a_{k}}{2},\frac{a_{k}+a_{k+1}}{2})\bigg]
\label{eq:disc-gaussian-sample}\end{align}

In this way, any Gaussian policy can be transformed into a trainable discrete policy, as demonstrated in Figure \ref{fig:actionspace} (c). For readability, for a continuous policy  $\pi(\cdot|s)=\mathcal{N}(\mu(s), \sigma(s))$, we write the discretized policy as $\textrm{Discretize}(\mathcal{N}(\mu(s), \sigma(s)))$. 
\subsubsection{Gaussian-and-Softmax Policy}
A straight-forward simplification of the discretized Gaussian policy is to replace the sampling of Eq.(\ref{eq:disc-gaussian-sample}) with a deterministic density function. So we can directly use a series of unnormalized logits $\bm{l}=(l_1, \dots, l_m)$ to get a Softmax policy, 
\begin{align}
    \apct_t|s_t \sim \textrm{Categorical}\big\{\textrm{Softmax}(\bm{l})\big\}.
\end{align}
where $l_k = -\frac{(a_k-\mu)^2}{2\sigma^2}$.
We write $\textrm{GSoftmax}(\mathcal{N}(\mu(s), \sigma(s)))$ to denote such a Gaussian-and-Softmax policy.
\subsubsection{HALOP with Two Stages} From the derivation, now we can get a discretized policy from any continuous Gaussian policy. It motivates us to feel free to design a two-stage generalization-first-specialization-second method under the same policy family and network structure.

\textbf{Stage 1:} learn the most generalized knowledge without distinguishing different types of stocks. Therefore, we have the following configuration. Firstly, the input states in this stage only contains the standardized public state $s^\textrm{pub}_t$ (by preprocessing all the prices and volumes to the same scale for different stocks), without the private execution status that reflects the characteristics of stocks. Second, the agent outputs the discretized percentage changes of price movements according to the realizable limit prices in order that the policy learned can be smoothly connected with the second stage. Formally, let $p^\textrm{c}_t$ be the current price of a stock, and $\theta_1$ be the parameters of the policy network in stage-1,
\begin{align}
\small
\mathcal{A}_{\textrm{S1}}=&\bigg\{\apct \bigg| \apct=\frac{\aticks\cdot\textrm{TickSize}}{p^\textrm{c}_t}, \aticks \in \mathcal{A}^{\textrm{ticks}}\bigg\},\\
\apct_{\textrm{S1},t} \sim& \pi_{\textrm{S1}}(\cdot |s^\textrm{pub}_t) = \textrm{Discretize}(\mathcal{N}(\mu_{\theta_1}(s^\textrm{pub}_t), \sigma_{\theta_1}(s^\textrm{pub}_t))),\\
&\aticks_{\textrm{S1},t} = \apct_{\textrm{S1},t}\cdot p^\textrm{c}_t / \, \textrm{TickSize}.
\end{align}
So the resulted output of the policy is a discrete action $\aticks_{\textrm{S1},t}$ in the tick action space, which represents a general guess of the limit price. We would like to use this action to scope a subset of actions around it, as illustrated in Figure \ref{fig:actionspace}(d).

\textbf{Stage 2:} After generalization, now for the specialization. We use the action offered by the policy of stage 1 as a medium point and search a fine-grained action around it in stage 2. We predefine a fixed-length window to scope the action subset of stage 2, i.e., consider a window-size of $2K+1$, the action subset of stage 2 is
$
    \mathcal{A}_{\textrm{S2}}=\big\{-K, \dots, 0, \dots, K \big\}
$.
Therefore, as illustrated in Figure \ref{fig:actionspace}(e), for any stock with arbitrary price, the action-space in stage-2 is always $2K+1$ integers, encouraging the agent to focus on localized optimization.
The state in stage 2 $s_t$ includes the raw public state without standardization and the private states, which contains the absolute value of prices, volumes, and the agent's execution status, characterizing each single stock. For stabilized learning, we scale the prices and volumes by taking the logarithms of the absolute values. Formally, let $\theta_2$ denote the parameters of the policy network in stage-2, we have
\begin{align}
\small
a_{\textrm{S2},t} \sim \pi_{\textrm{S2}}(\cdot|s_t) &= \textrm{GSoftmax}(\mathcal{N}(\mu_{\theta_2}(s_t), \sigma_{\theta_2}(s_t)))\\
&\aticks_{\textrm{S2},t} = \aticks_{\textrm{S1},t} + a_{\textrm{S2},t}.
\end{align}

\begin{figure}[t]
    \centering
    \includegraphics[width=0.48\textwidth]{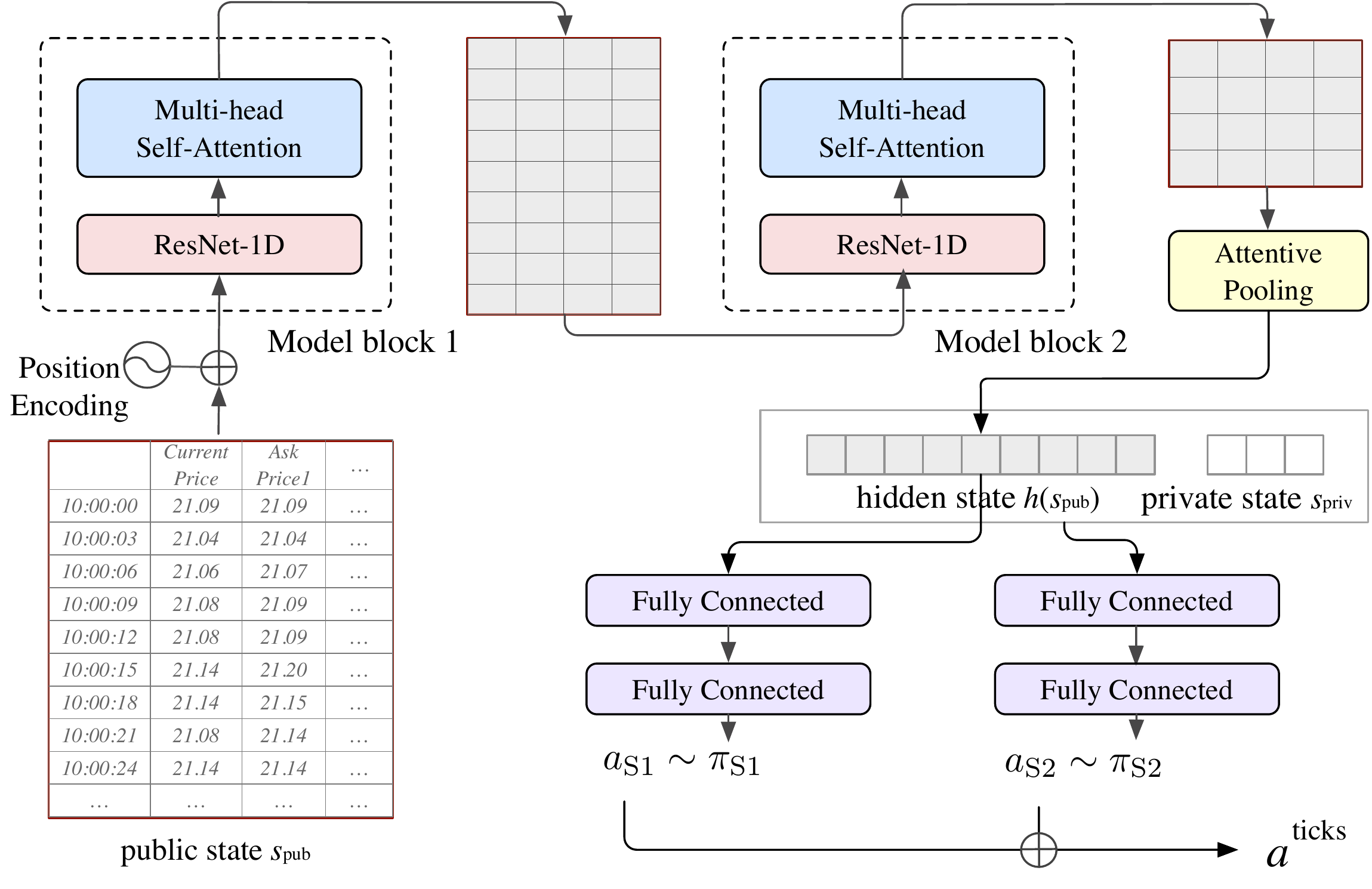}
    % \caption{The neural network architecture and feed-forward workflow of HALOP.}
    \caption{The architecture of HALOP.}
    \label{fig:net}
\end{figure}

\subsection{Architectures and Implementation Details}
In this part, we detail the choice of neural network architecture for learning from high-dimensional sequence data, as well as the learning details of PPO.

The overall network structure is shown in Figure \ref{fig:net}. The input of the representation learning network is a sequence of high-dimensional tick snapshot data, i.e., the public states, with tens of real-valued features and hundreds of timesteps. 
First, we use two stacked neural blocks for sequence encoding, each block consists of a residual network with an 1-D convolutional neural network (1DCNN) to capture temporal patterns and shorten the sequence length, and a multi-head self-attention layer \cite{vaswani2017attention} to learn the correlation among time steps.
Finally, we get a dense representation $h(s)$ for the state by attentive pooling the output of the self-attention matrix. 

% Following similar spirit of recent time-series sequence learning methods \cite{lai2018modeling,zhou2021informer}, we first use a residual network with 1-D convolutional neural network blocks to capture temporal patterns while shortening the sequence length. Then, we apply 2 self-attention \redfont{add citation} layers to learn the correlation between the features of the extracted time steps. Finally, we apply an average pooling to the output of the self-attention matrix to get a flatten dense representation $\bm{h}(s)$ for the state.

Second, we use an actor-critic architecture \cite{DBLP:journals/corr/SchulmanWDRK17,pan2019pome} for the policy optimization. For both stage-1 and stage-2, we use two fully-connected layers as the policy and value networks. In stage-1, for generalization, the networks directly take the public $h(s^\textrm{pub}_t)$ as input. In stage-2, we concatenate the private state to $h(s^\textrm{pub}_t)$ to get a new dense vector as input. The two stages share the same representation learning network for better sample efficiency.

For policy optimization, we regard the order execution process of each (trading day, stock) pair as a single \emph{episode}, which has $T$ steps. Then, we group the episodes of all the stocks in each trading day as an $epoch$. For example, if there are 250 trade days and 300 stocks, then it groups into 250 epochs and $250\times 300$ episodes. During training, we iterates between rollouts and updates in a batch RL manner. At each round, we randomly select one trading day from the training set, and run parallel rollout for all the episodes in that trading day. After finished, we perform the on-policy policy update with the collected data like the standard PPO algorithm \cite{DBLP:journals/corr/SchulmanWDRK17}. 

\begin{table*}[t]\small
\centering
\label{tab:mainres}
\begin{tabular}{l|cccc|cccc|cccc} \toprule
\multirow{2}{*}{Method} & \multicolumn{4}{c|}{TWAP} & \multicolumn{4}{c|}{VWAP} & \multicolumn{4}{c}{ODP} \\ \cmidrule{2-13} 
 & Return & PnL & Std & $t$-value & Return & PnL & Std & $t$-value & Return & PnL & Std & $t$-value \\ \midrule
Market Order & 0.00 & 0.00 & 0.00  & 0.0  & 0.14 & 0.14 & 8.72  & 3.1  & 1.64 & 0.92 & 65.34 & 4.8  \\ \midrule 
PPO-Gaussian & 1.41 & 1.33 & 4.61  & 58.7 & 1.38 & 1.31 & 8.94  & 29.7 & 2.20 & 1.12 & 64.55 & 6.5 \\ 
PPO-Softmax &  2.12 & -1.46 & 35.88 &  11.3 &  2.36 & -1.17 & 35.47 &  12.8 & 3.34 & -0.53 & 66.73 &   9.6 \\ \midrule
HALOP Stage-1 & 3.69 &  1.46 & 17.05 &  41.5 & 3.87 &  1.68 & 17.51 &  42.4 & 4.51 &  1.45 & 63.67 &  13.6 \\
HALOP & {\bf 4.41} & {\bf 2.98} & 6.87  & 123.2 & {\bf 4.52} & {\bf 3.12} & 9.24  & 93.8 & {\bf 5.04} & {\bf 2.46} & 63.42 & 15.3  \\ \bottomrule
\end{tabular}
\caption{Main results.}
\end{table*}

\section{Experiments}
Our experiments are designed to investigate the following questions: 

\textbf{Q1.} Can HALOP beat the market? 

\textbf{Q2.} Can HALOP improve upon other volume-oriented order splitting strategies? Is HALOP stable and robust?

\textbf{Q3.} What benefits can we get from HALOP compared with vanilla continuous or discrete control? 

\textbf{Q4.} Can HALOP address the trade-off between generalization and specialization better than other methods?
% 最后一点想说明可以加volume，且同样有效

\subsection{Experiment Setup}

\subsubsection{The Simulation Environment} We conduct experiments in simulation environments which is built over historical high-frequency transaction data of stocks of the CSI 300 index in the China A-shares. Specifically, we test all the methods in the buying direction and directly use the raw market information as the agent's state, which consists of the top 5 bid/ask prices and volumes, and the current (last) price. 
As described in the previous section, we use the data from January 2010 to December 2019 to build the training episodes, and use the data from January 2020 to June 2020 to build the testing episodes. 

At each episode (i.e., for one stock on one trading day), our agent's execution mission starts at 10:00:00 (half an hour after opening) and ends at 14:30:00 (half an hour before close) which is an 180-minute trading hour. The time period is then split into $T=90$ intervals evenly. At each step, the agents observes the market information and send an order to the market. We simulate a three second communication delay so as to make the experiment close to reality. For example, an order sent at 10:02:00 
can be executed after 10:02:03. Moreover, to avoid potential large market impact, our environment does not execute any large order whose size is larger than $1/10$ of the agent's total inventory of the day. So on receiving an order with size $v_t$, the environment executes no more than $\min(v_t, 1/10)$ and cancels the rest of it, if there is. %Notice such ``cancellation'' also occurs for market orders in our simulation.
% Therefore, even when using market orders, such ``cancellation'' still occurs in our simulation.

\subsubsection{Evaluation Metrics}
Our evaluation metrics are designed around two core concepts in quantitative finance: the excess return and the risk.

We use the average excess return over TWAP-with-market-order as the \textbf{Return} metric. That is, the TWAP strategy using market orders always get a return of 0, and other methods get positive returns if it beats TWAP. The returns are displayed in basis points (bps, 1 bp equals to $0.01\%$). 

Next, we compute the standard deviation of excess returns as a risk indicator, denoted as \textbf{Std}. Smaller standard deviation of returns means that the algorithm is more stable throughout different time and stocks.

With the computed Return and Std metrics, we calculate the {\bf $t$-value} for paired student's $t$-test, to show the significance of the excess return. For any method, given the metric Return and Std, the $t$-value can be computed by 
\begin{equation}
    t\textrm{-value} = \frac{\textrm{Return}}{\textrm{Std} / \sqrt{n-1}},
\end{equation}
where $n$ is the number of episodes. A larger $t$-value indicates more significant results. When $n$ is large enough, $t > 3.29$ means that there is a improvement over the benchmark in confidence level 99.9\%.

Finally, we need a more realistic metric that takes both return and safety constaints into consideration, which is our \textbf{PnL} (or Profit and Loss) for order execution. The profit is the excess return. For the loss, we take a regulation constraint into consideration: the cancellation rate for institutional investors cannot be greater than 50\%. So when the cancellation rate exceeds 50\%, we penalize the agent by -5 bps. So
\begin{equation}
    \textrm{PnL} = \textrm{Return} - \sum_i^n 5\cdot\mathbb{I}\big[\Delta_T + \sum_{t=1}^T v_t > 2\big],
\end{equation}
with $v_t$ defined in Eq.(\ref{eq:vt-def}) and $\Delta_T$ the size of market order at the end of an episode.
\subsection{Compared Methods}
There are two categories of methods we compared: the price- and volume-oriented optimization methods. For the first part, we compare the following methods:
\begin{itemize}
    \item \textbf{Market Order} which place market orders at all steps, meant to be executed as quickly as possible.
    \item \textbf{PPO-Gaussian} which is PPO \cite{DBLP:journals/corr/SchulmanWDRK17} with a Gaussian policy (Figure \ref{fig:actionspace}(a)). 
    \item \textbf{PPO-Softmax} which is PPO \cite{DBLP:journals/corr/SchulmanWDRK17} with a Softmax policy (Figure \ref{fig:actionspace}(b)). 
    % of the  Gaussian policy as in Figure \ref{fig:actionspace} (a).
    \item \textbf{HALOP} is our proposed method with two stages. 
    \item \textbf{HALOP Stage-1} is HALOP with only one stage (using the discretized Gaussian policy). It takes both public and private states as inputs. We test it to see whether it can improve upon the ordinary Gaussian policy.
\end{itemize}
For a fair comparison, all the RL algorithms are tested with the same network architecture (shown in Figure \ref{fig:net}). The number of actions in HALOP stage-2 is set to $7=2\times3+1$.

To see how the agents perform under different volume schedule, we test the following volume-oriented methods:
\begin{itemize}
    \item \textbf{TWAP} which evenly splits the order to $T$ pieces and execute the same amount of shares at each step. 
    \item \textbf{VWAP} which sends orders in proportion to an estimated volume ratio of each time period of the day, which is estimated from the previous 21 trading days.
    \item \textbf{OPD} \cite{fang2021universal} which leverages RL with policy distillation to determine the size of each sub-order.
    % \item \textbf{discrete-PPO} utilizes PPO algorithm with action space of the discrete softmax policy as in Fig.(b).
\end{itemize}

\subsection{Main Experimental Results}

Table \ref{tab:mainres} reports the overall performance.

First, to answer question \textbf{Q1}, it is clear that our HALOP method can indeed outperform the market orders with a large margin. When using the basic TWAP, HALOP beats Market Order with 4.41 bps and a PnL of 2.98, with a $t$-value of 123.2 which proves the significance of improvements.

Second, to answer \textbf{Q2}, we would like to see how HALOP performs under other volume allocation methods. We firstly see that the superiority of HALOP againest Market Order still holds with either TWAP, VWAP or ODP. Also, surprisingly, we find that when using HALOP, the Std of ODP decreased comparing to the original ODP with market order, indicating that HALOP can even stabilize the order execution process in certain circumstances. 

Third, to answer \textbf{Q3}, we need to compare HALOP with ordinary PPOs as well as the lite version of HALOP with only stage-1. We find that though PPO-Gaussian and PPO-Softmax all yields positive excess returns, the improvements are not as large as that of HALOP. Moreover, while PPO-Softmax gets a higher excess-return than PPO-Gaussian, it has negative PnL and very large Std. It indicates that although the Softmax policy is capable of finding the best action, it is unstable and unreliable. We also see that HALOP Stage-1 yields significantly better result than the ordinary PPOs, showing its ability of taking the advantages of both Gaussian policy and discrete control. Finally, the comparison between HALOP and HALOP Stage-1 shows that the specialization in stage-2 indeed helps improve both excess return and stability.

\subsection{Grouping Study}
We further design a grouping study to see whether HALOP address the trade-off between generalization and specialization well. We group the test episodes by stock close prices:
``low-priced'' for price less than 10.00 CNY, ``medium-priced'' for price between 10.00-50.00 CNY, and ``high-priced'' for larger than 50.00 CNY.
\begin{figure}
    \centering
    \includegraphics[width=0.78\linewidth]{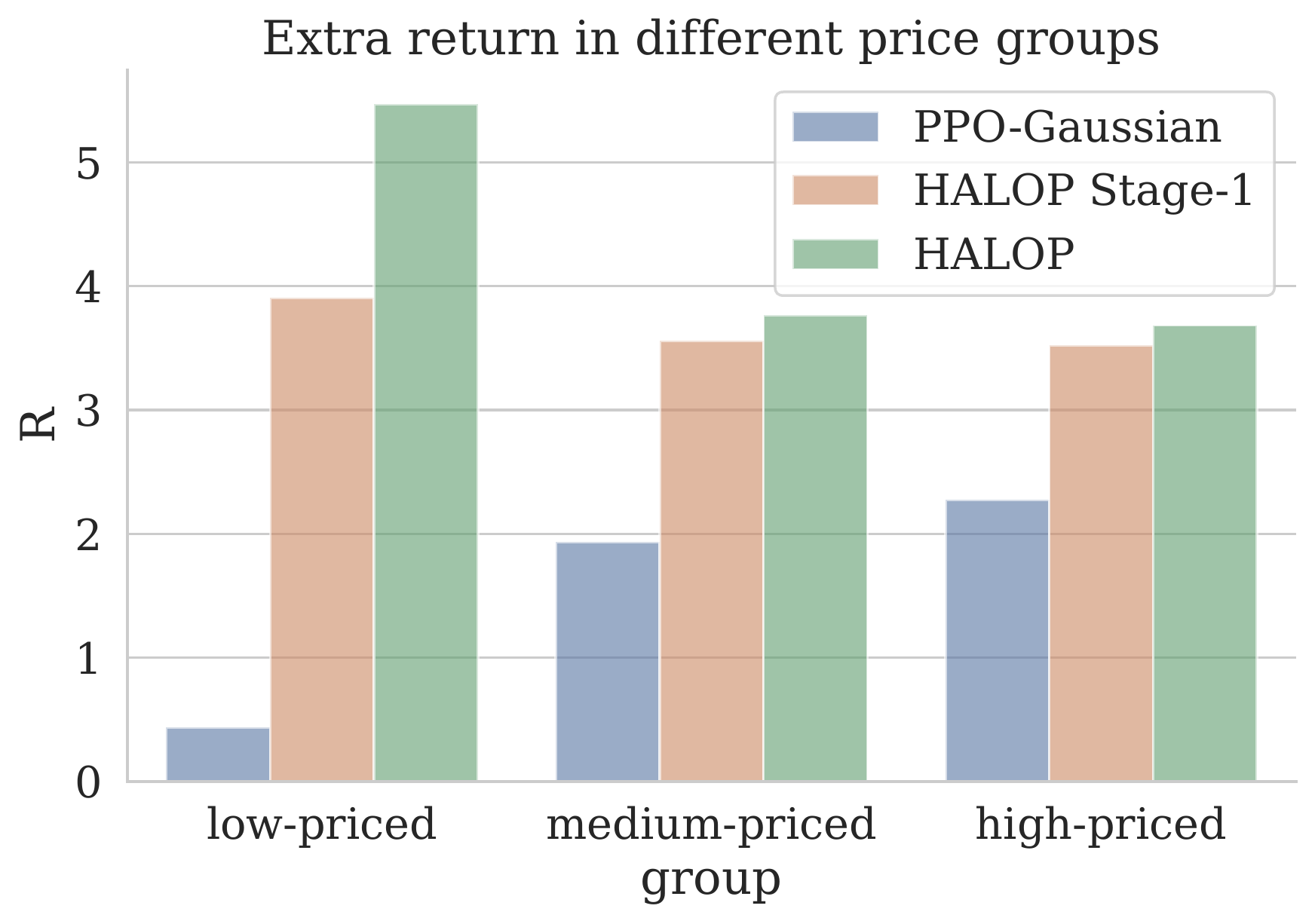}
    \caption{Results of grouping study}
    \label{fig:grouping}
\end{figure}

As shown in Figure \ref{fig:grouping}, we find PPO-Gaussian more in favor of high-priced stocks than low-priced stocks. Next, we observe that HALOP Stage-1 has a strong ability of generalization in all the groups: the excess returns are all over 3.5. Further, the two-stage HALOP further improves the results especially for low-priced stocks, which indicates that the fine-grained discrete control in stage-2 indeed achieves better specialization for different types of stocks. Specifically, for penny stocks which are known to have less liquidity, the discretization in HALOP Stage-2 is shown to be more beneficial.

\section{Conclusion}
This paper focuses on optimal execution with limit orders by setting better limit prices. We propose a novel reinforcement solution, Hybrid Action-Space Order Placement, to address the generalization-specialization trade-off by combining both advantages of continuous and discrete control. The proposed method has two stages: in stage-1, the agent selects a target percentage change in price with a discretized Gaussian policy and scopes a subset of discrete actions thereby; in stage-2, the agent uses a fine-grained Gaussian-and-Softmax policy to select a tick-based action. Extensive simulation-based experiments show that our method can beat the market and improve upon other volume-oriented order splitting strategies. It shows stable and robust improvements comparing with vanilla continuous or discrete control methods. Specifically, the discretized Gaussian policy of stage-1 helps improve the generalization ability, and meanwhile in stage-2 the agent specializes in policy for different types of stocks especially for low-priced stocks. 
In a broader sense, we believe that HALOP offers a reference of applying RL in real-world applications where the trade-off between generalization and specialization should be taken into consideration.

\bibliographystyle{named}
\bibliography{ijcai22}

\end{document}